\documentclass[12pt]{article}

\usepackage{amsmath, setspace}
\usepackage{amssymb, xspace}

\usepackage{graphicx}
\usepackage{color} 

\usepackage{subcaption}
\usepackage{hyperref}
\usepackage{times}
\usepackage[backend=biber,style=nature]{biblatex}

\usepackage[table,rgb]{xcolor}
\usepackage{multirow}
\usepackage[misc,geometry]{ifsym}

\usepackage{caption}
\DeclareCaptionLabelSeparator{pipe}{$\; | \;$}
\definecolor{abstarctBlue}{rgb}{0.0706, 0.349, 0.6667} 
\definecolor{abstarctBlue2}{cmyk}{ 0.5143,   0.2857,    0.0286,    0.3714}
\definecolor{Section}{cmyk}{0.2,0.8,0.8,0.3}

\usepackage{soul}
\sodef\an{\fontfamily{phv}\selectfont}{.08em}{1em plus1em}{0.5em plus.1em minus.1em} 
\sodef\ann{\fontfamily{phv}\selectfont}{0.04em}{0.5em plus0.02em}{0.1em plus.1em minus.1em}

\makeatletter
\renewcommand{\@biblabel}[1]{\quad#1.}
\makeatother

\date{}

\usepackage{overpic}
\newcommand*{\hvfont}{\fontfamily{phv}\selectfont}

\newcommand{\inb}[2]{ \begin{overpic}[width = .42\textwidth]{#1} \put(0,75){\large \bf \hvfont #2}\end{overpic} }

\newcommand{\fg}{\textcolor{linkcolor}{Fig.}~\ref}
\newcommand{\supp}{Methods\xspace	} 

\newcommand{\ith}{\ensuremath{i^{th} }\xspace	} 
\newcommand{\jth}{\ensuremath{j^{th} }\xspace	}

\newcommand{\pana}{({\bf a})\xspace}
\newcommand{\panb}{({\bf b})\xspace}
\newcommand{\panc}{({\bf c})\xspace}
\newcommand{\pand}{({\bf d})\xspace}
\newcommand{\pane}{({\bf e})\xspace}
\newcommand{\panf}{({\bf f})\xspace}

\definecolor{citecolor}{rgb}{0.071, 0.36, 0.67}   
\definecolor{linkcolor}{rgb}{0.071, 0.4, 0.67}  
\hypersetup{
colorlinks=true, 
citecolor=citecolor,
linkcolor=linkcolor,
urlcolor=linkcolor
}

\newcommand{\name}{SCoPE-MS\xspace	} 

\graphicspath{ 
	{Figs/}	   
}

\let\citep=\autocite
\let\citet=\autocite  

\def \simpson { \footnotesize {\bf \hvfont Simpson's paradox confounds the interpretation of population-average protein and mRNA measurements} \hvfont
\pana For a particular gene, its protein levels across tissues can be poorly predicted by its mRNA levels, while  the average protein levels can be well predicted by scaled mRNA levels \citep{Franks2016PTR}. Thus mRNAs levels are unreliable surrogates for relative protein levels, and we need direct measurements of proteins.  \panb A related manifestation of Simpson's paradox indicates that the average levels of the \ith and the \jth proteins may appear positively correlated, even though they are inversely correlated within a cell-type. Averaging across cells, even cell types sorted based on markers, will obscure the the relationship between the \ith and the \jth proteins.   
}

\def \opportunities  { \footnotesize {\bf \hvfont Transformative opportunities for improving the quantification of single-cell proteomes.} \hvfont
\pana Most bulk samples prepared for MS have volume of $10-100\; \mu l$\citep{aebersold2003mass, zhang2014proteogenomic,Slavov_exp}. 
 Reducing the volume for sample preparation $100\times$ to $nl$ can significantly reduce protein losses from the walls of tubes. 
\panb The sharper the separation peaks, the larger fraction of the ions can be analyzed for a fixed sampling (injection) time. Sharper peaks can be achieved  by reducing the bore of LC columns, using monolithic columns, PLOT columns\citep{Ivanov2015RareCells}, or capillary electrophoresis\citep{lombard2016single}. 
\panc Typically elution peaks have a full width at the base of about $60\; s$ and about $10-15\; s$ at mid-height, while ions for MS2 are sampled for mere milliseconds. These settings are typical for bulk proteomics and result in sampling less than $1\; \%$ of the ions delivered to the instruments. Thus, increasing the sampling time $100\times$ can substantially increase the ions analyzed by MS, the sensitivity, and the accuracy of quantification.   While the panel displays sampling during the apex of the peak, this cannot always be achieved for all ions.    
\pand Automated liquid handling and 96 / 384 well plates can increase the consistency of sample preparation, decrease volumes to the $nl$ range and increase throughput.  \pane Parallel accumulation and serial injection of ions can afford increased ion sampling without reducing throughput. \panf Larger number of barcodes will increase the number cellular proteomes quantified per run without reducing proteome coverage or ion sampling.      
 }
 
\date{}
\begin{document}

\begin{spacing}{1.6}
\noindent {\LARGE \bf 
Routinely quantifying single cell proteomes:\\A new age in quantitative biology and medicine
}
\end{spacing}

\vspace{10mm}

\noindent\ann{\large
Harrison Specht$^{1}$ 
\& 
Nikolai Slavov$^{1,2,}$\textsuperscript{\Letter }
} 

{
\noindent 
$^{1}$Department of Bioengineering, Northeastern University, Boston, MA 02115, USA\\
$^{2}$Department of Biology, Northeastern University, Boston, MA 02115, USA\\
{\Letter} Correspondence: {\an{\small nslavov@alum.mit.edu}}
}\\

\thispagestyle{empty}


\begin{spacing}{1.2} 
\noindent{\hvfont 
Many pressing medical challenges -- such as diagnosing disease, enhancing directed stem cell differentiation, and classifying cancers -- have long been hindered by limitations in our ability to quantify proteins in single cells. Mass-spectrometry (MS) is poised to transcend these limitations by developing powerful methods to routinely quantify thousands of proteins and proteoforms across many thousands of single cells. We outline specific technological developments and ideas that can increase the sensitivity and throughput of single cell MS by orders of magnitude and usher in this new age. These advances will transform medicine and ultimately contribute to understanding biological systems on an entirely new level.                
}
\end{spacing}
\vspace{-8mm}

\begin{spacing}{1.3} 
\section*{}
Quantifying proteins in single cells has a long history. For decades, scientists and physicians have used antibodies, fluorescent proteins, and MALDI-TOF to identify or quantify a few different proteins per cell \citep{chalfie1994green, caprioli1997molecular, bendall2011single, darmanis2016simultaneous, hughes2014single}. These methods have enabled new discoveries\citep{elowitz2002stochastic}, clinical applications\citep{weissleder2008imaging}, and even spawned new fields, such as understanding the role of noise in gene expression \citep{ raj2008nature}. These impressive achievements were made based on measuring just a few different proteins per cell.\\

\noindent
However, many pressing needs in medicine, such as diagnosing disease and enhancing directed stem cell differentiation, as well as transformative opportunities in biology, demand qualifying $100 - 1000$ fold more proteins; they demand an entirely different set of approaches and techniques. Such approaches are beginning to coalesce around new ideas and emerging technologies in MS-based proteomics that promise quantifying thousands of proteins and all their modifications (termed proteoforms) across thousands of single mammalian cells. We begin by outlining the urgent demand and exciting opportunities for these methods since this context can best motivate future methodological developments. Then we discuss specific technical opportunities that can increase the sensitivity and throughput of single cell proteomics by orders of magnitude and thus contribute to realizing its tremendous promise.

\subsection*{Limitations of population-average measurements} 
The ability to quantify nearly complete transcriptomes and proteomes from bulk samples has been transformative for biomedical research \citep{brown1999exploring, aebersold2003mass, zhang2014proteogenomic}. It has enabled unbiased screens and unexpected discoveries \citep{boyer2005core,Slavov_batch_ymc,Slavov_exp}. Yet  comprehensive proteome measurements have been confined to samples comprised of  many cells, reflecting only the population-average. Interpreting population-average protein levels is fundamentally confounded when samples consist of heterogeneous cells. The most obvious caveat is that the population-average may not be representative for any cell. For example, proteoforms may have bimodally-distributed abundances within the whole heterogenous cell population. 
More generally,  
trends within groups -- within different cell types, for instance -- may disappear or even reverse when these groups are combined, as with population-average measurements. This phenomenon is known as Simpson's paradox\citep{blyth1972simpson}. We recently demonstrated\citep{Franks2016PTR} the confounding effects of Simpson's paradox in using mRNA levels as surrogates for protein levels with bulk data, \fg{simpson}a. Similarly, Simpson's paradox can confound the interpretation of population-average protein levels as illustrated in \fg{simpson}b. If we only consider the average levels of the \ith and the \jth proteins across cell types,  the proteins seem positively correlated  (\fg{simpson}b). However, paradoxically, within each cell type the abundances of the \ith and the \jth proteins can have an inverse relationship (\fg{simpson}b). Perfect measurements of cell types sorted based on a few markers cannot resolve such phenomena; the true relationship can be observed only by measuring proteins in single cells.  \\

\begin{figure}[h!]
\centering
   \inb{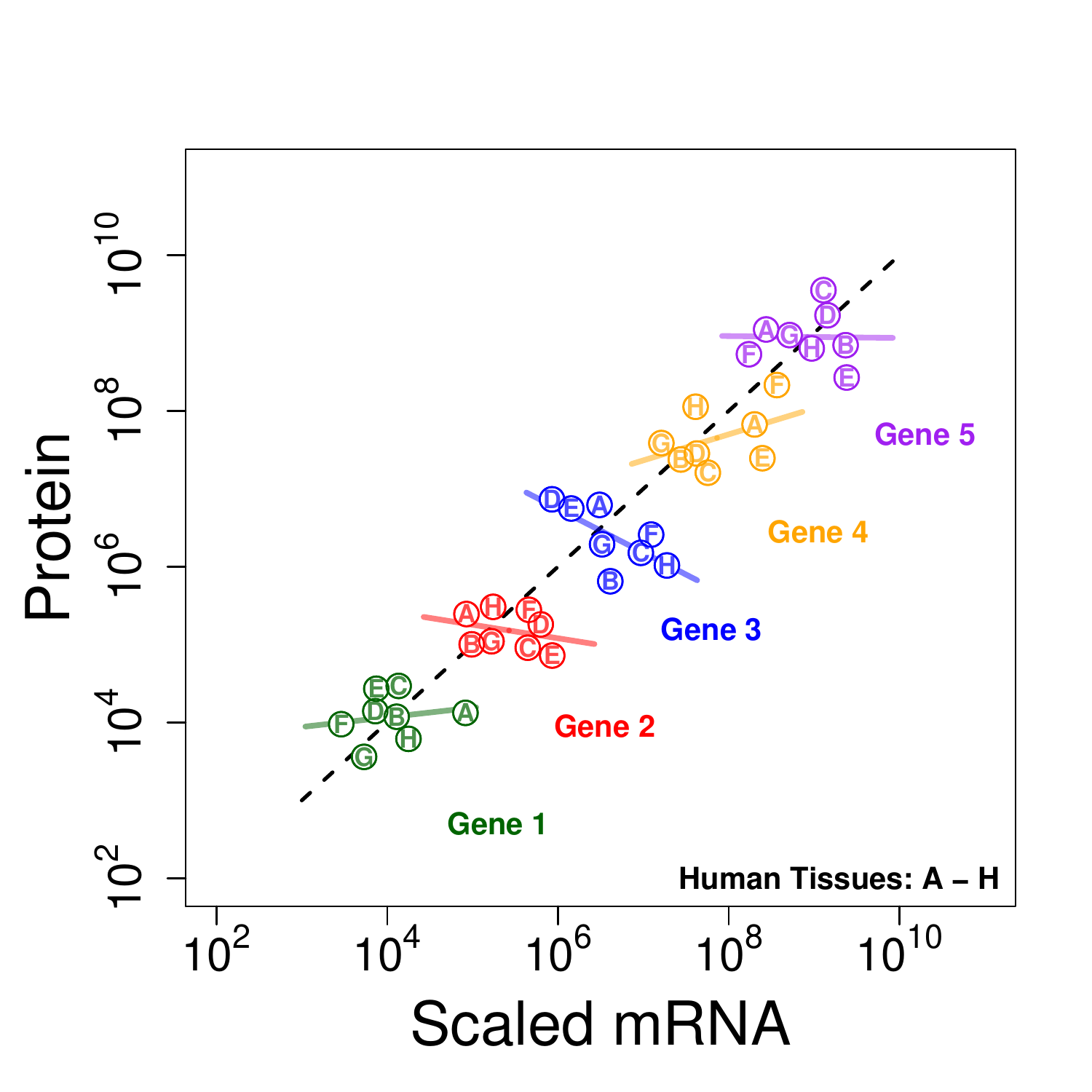}{a}
    \hspace{0.08\textwidth}
   	 \inb{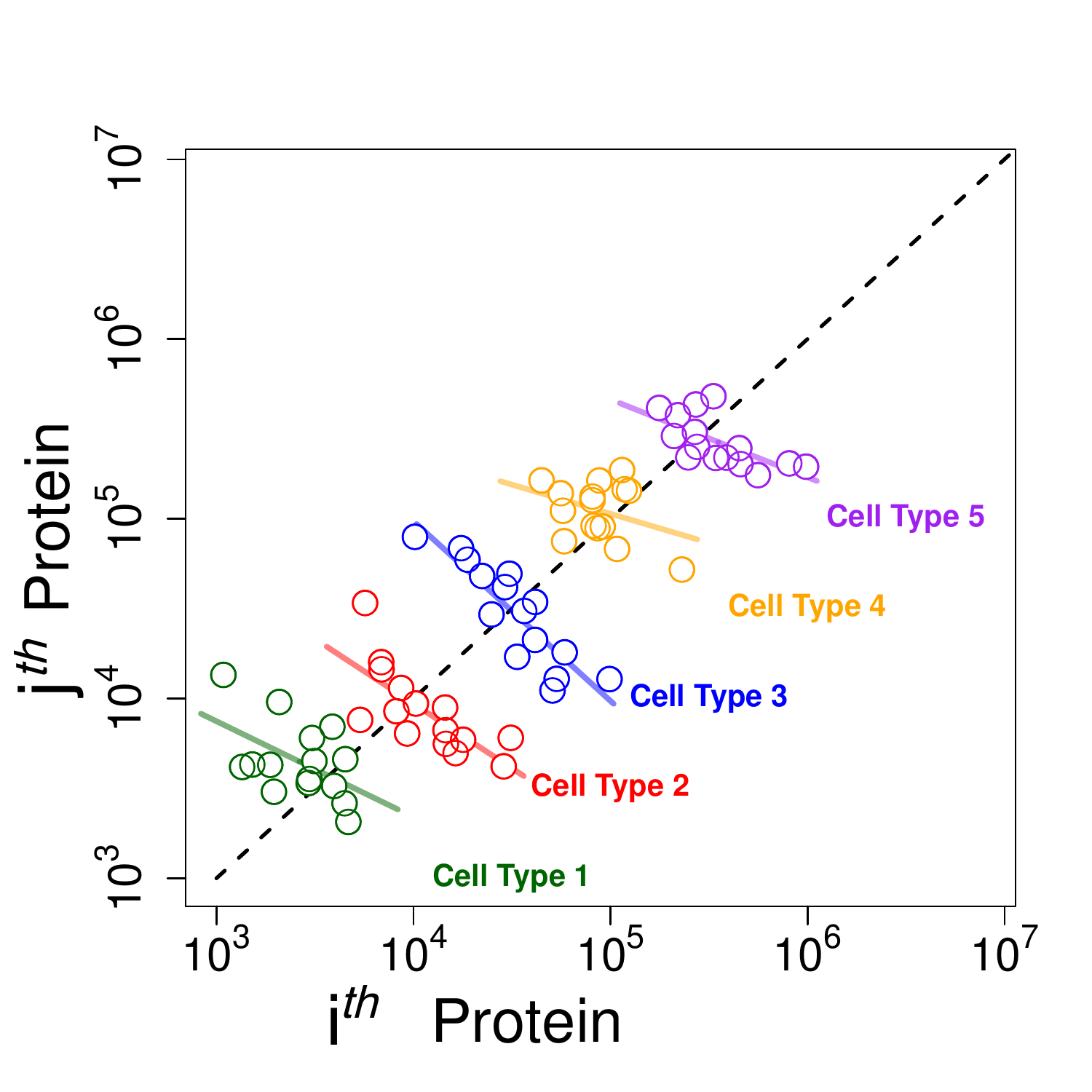}{b}
	\caption{\simpson}
	\label{simpson}
\end{figure}

\noindent
For the reasons illustrated in the examples above, measuring protein levels in tissues -- both for establishing healthy baselines and for diagnosing disease -- is best done by measuring protein abundance in single cells. The most immediate clinical applications of single cell proteomics are the discovery of biomarkers and their diagnostic use.  In the longer term, single cell proteomics can open new frontiers in unbiased modeling, understanding, and rational control of biological systems as we discuss below.

\subsection*{Single-cell proteomics opens exciting frontiers}
Single cell proteomics will allow us to perform causal protein inference and create unbiased models of direct and indirect protein interactions. These exciting prospects require estimating joint and conditional probability distributions of proteins across single cells, which in turn require many observations of individual proteins quantified across thousands of individual single cells. Such estimates have been obtained only for a few proteins, relying on antibodies or fluorescent proteins. Reliably-estimated joint distributions of proteoforms enable modeling protein-protein dependencies with the empirical probabilities, without assuming a sigmoidal, linear, or any specific relationship. Such models can infer the dependence and its causality between any two measured proteoforms while controlling -- without assumptions -- for the influence of all other measured proteoforms. Yet, these models can control only for the measured proteins. Thus, the full power of such analysis, which can lead to causal inference, requires measuring all relevant proteoforms.  We believe that high-throughput single-cell proteomics can estimate the joint distributions among the relevant proteforms, and in the process catalyze a transition from the population-average measurements to casual inference, elucidating direct and indirect protein interactions and signaling mechanism.  Similarly, the joint distributions enable estimating the mutual information (MI) between proteins. The MI inequality states that the mutual information between a variable $X$, e.g., a kinase, and its causal variable, e.g., an upstream kinase, is always larger than the  mutual information between $X$ and another correlated but not causal variable. This allows us to determine which protein kinases are upstream and downstream in a signaling cascade, or another similar biological sequence of causal events. Such progress will produce models that can predict and design the outcomes of new treatments, e.g, rationally engineer directed differentiation of stem cells into cell types of interest. 
Realizing these promises requires increasing both the sensitivity and the throughput of of single-cell proteomics. We believe both factors can be increased by orders of magnitude based on the suggestions outlines below. \\

\section*{Transformative opportunities for realizing single cell proteomics}
In the foreseeable future, two types of technologies are likely to increase the number of proteins quantified per single cell: $(i)$ antibodies-based methods and $(ii)$ MS-based methods. Antibody-based methods have a  long track-record of success. For many years, the binding of antibodies to a few cellular proteins have been assayed by measuring fluorophores, transition metals and more recently DNA sequences conjugated to the antibodies. Such methods have measured up to a few dozen proteins per cell. Efforts are underway to increase this number while ensuring cellular permeability, antibody availability, binding specificity, epitope availability, and overcoming molecular crowding limitations.    
In contrast to antibody-based methods, quantifying proteins in single cells by MS is in its infancy.  Yet, initial attempts have already quantified hundreds of proteins per cell and thousands of proteins over many single cells\citep{scopems2017}.  Below we focus on MS-based methods as we believe they have the potential to afford unparalleled specificity, measurement accuracy, depth of proteome coverage and flexibility in experimental design.

So far, most single cell MS studies have used either $(i)$ MALDI-TOF\citep{caprioli1997molecular} whose quantitative accuracy is undermined by variable and incomplete ionization  or $(ii)$ quantitative electrospray ionization MS methods applied to unusually large cells, i.e., human oocytes \citep{oocytes2016Krijgsveld}  or frog embryos \citep{lombard2016single} that exceed the size of a typical mammalian cell ($\sim 15 \mu m$ diameter) by orders of magnitude \citep{milo2010bionumbers}. Protein measurements in typical mammalian cells have quantified thousands of proteins in cell lysates corresponding to hundreds of cells \citep{Ivanov2015RareCells}, and just recently we reported a method that allows quantifying over a thousand proteins across many single mouse stem cells\citep{scopems2017}. We believe that the field is ready to take off from this launching point, and increase the number of accurately-quantified proteins and the number of single cells assayed by orders of magnitude.

The relatively high copy number of proteins per cell\citep{milo2010bionumbers} can support both deep proteome quantification and low sampling error compared to  single cell RNA-seq. So far, RNA-seq has quantified the most molecules per cell but the accuracy of these measurements has been limited by counting noise: since only a subset of the RNA molecules are sampled (counted), their estimated abundances contain missing values and counting error due to sampling\citep{hicks2017missing}.  If we count $n$ molecules out of a larger pool, the Poisson distribution estimates the standard deviation of the sampling as $\sqrt n$, and thus the relative error, estimated as standard deviation over mean, is  $\sqrt n / n = 1/\sqrt n$. Therefore, if we sample $20\%$ of the mRNA molecules from a relatively abundant gene and obtain $4$ sequences per cell, we can expect about $1/\sqrt{4} = 50\%$ relative error from the sampling alone. Since cells contain over a $1000$ protein molecules per mRNA molecule\citep{milo2010bionumbers}, the counting error of protein sampling is likely to be smaller: For the median protein, having $50,000$ molecules per cell, sampling even $1\%$ of the molecules will result in about $1/\sqrt{500} \approx 5\%$ sampling error. 
 These numbers promise deep and accurate quantification of single-cell proteomics if we can accomplish the opportunities discussed below. The challenges are technical -- posing no conceptual limitations -- and we believe they are tractable with current and emerging ideas and technologies.  Many of these challenges and opportunities are similar for bottom-up methods (measuring peptides after protease digestion) and for top-down methods (measuring undigested proteins) and we discuss them together, although top-down proteomics faces lower sensitivity, harder multiplexing and more challenging protein identification.

\begin{figure}[h!]
	\includegraphics[width=0.98\textwidth]{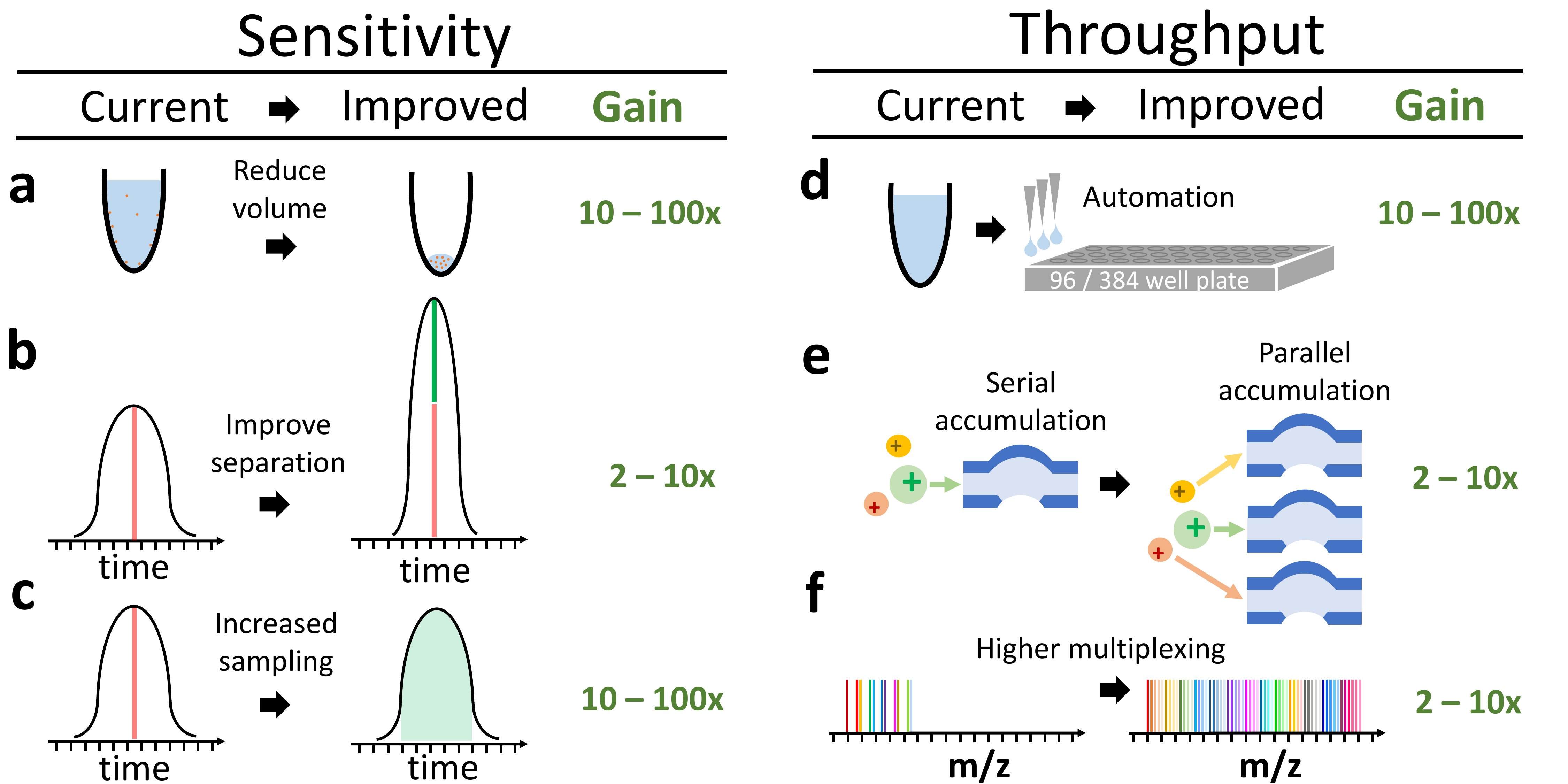}\\[1em]
	\caption{\opportunities}
	\label{opportunities}
\end{figure}

\subsection*{Sample preparation}
Cell extraction and separation for single cell proteomics can be similar to cell extraction for other analytical methods, i.e., single cell RNA-seq, and such methods have been reviewed extensively\citep{hosic2015microfluidic}. After extraction and separation, sample preparation for single cell proteomics  should ideally result in complete cell lysis and digestion, incur minimal loss of protein, and allow high-throughput automation for the analysis of thousands, even millions, of single cells. Standard MS sample preparation methods use $10-100 \; \mu l$  volumes; reducing these volumes to  $1\; \mu l$ or less will afford a substantial reduction in protein loss and reagents used, \fg{opportunities}a.  This protein loss is primarily due to surface adsorption, which bulk proteomics overcomes by operating near the protein or peptide solubility limit.  To mimic that approach for single cell proteomics, given the $ \sim 500\;  pg$ of protein in a typical mammalian cell \citep{milo2010bionumbers}, the reaction volume should be limited to nanoliters. While techniques for surface passivation permit that volume to expand, automated precision liquid handling will play a large role in single cell proteomics, making sample preparation robust, high-throughput, and cost-effective at scale. Furthermore, cell lysis methods that use only MS-compatible reagents, e.g., sonication in water \citep{Ivanov2015RareCells, scopems2017},  can obviate cleanup and further reduce losses. Such approaches can increase sensitivity and throughput by orders of magnitude,  \fg{opportunities}a,d.  

\subsection*{Peptide separation} 
Modern MS methods sample only a fraction of each elution peak, which corresponds to 1\% or less of the ions for each peptide that the instrument could sample (\fg{opportunities}b). This small fraction is sufficient  for the identification of peptides in bulk samples and maximizes peptide identification per unit time. Single cell proteomics, however, can benefit from delivering the maximum number of ions from each peptide to the MS detectors. Two simple strategies can maximize the number of ions delivered: ($i$) improved peptide separation with tighter, and therefore taller, peaks, so that the number of ions sampled per unit time is increased; ($ii$) sampling a greater fraction of the elution peaks. The first strategy is technically demanding while the second one comes at the expense of reduced throughput per unit time. Such limitations can be, at least partially, compensated by ($i$) improving the separation, i.e., using specialized liquid chromatography (LC) columns\citep{Ivanov2015RareCells} or capillary electrophoresis \citep{lombard2016single} (which is particularly promising for top-down MS), or by ($ii$) making all measurements targeted to proteins of interest, which involves programming the MS instrument to sequence specific peptides. These strategies, alone or combined, have the potential to increase the sensitivity of single cell proteomics by $10 - 100$ fold, \fg{opportunities}b,c.

\subsection*{Parallel ion accumulation}
Data Dependent Acquisition (DDA) with current MS hardware operates by selecting specific ions and fragmenting them for identification and quantification. The ions are serially accumulated and serially analyzed. This strategy is efficient when accumulation times are short. In bulk samples, ions are abundant enough so that sampling them for a very short time results in reliable identification and quantification. However,  single-cell MS ideally should sample a large fraction of the ions, which requires accumulating ions over longer periods of time. If ion accumulation is serial, longer accumulation increases sensitivity but at the expense of fewer quantified peptides per unit time. 
This trade-off can be resolved by accumulating ions in parallel, so that we gain sensitivity without losing throughput. The most obvious way to accomplish parallel accumulation and parallel injections is by using Data Independent Acquisition (DIA). Since DIA does not allow direct quantification from the reporter ions of isobaric tags, we can adapt it to use the complement ions that remain bound to peptide fragments. Parallel accumulation with serial injection can be accomplished by using additional accumulation traps built into the MS instrument that allow parallel accumulation but serial injection (\fg{opportunities}e). Alternatively, parallel accumulation and serial injection can be achieved by trapped ion mobility spectrometry\citep{meier2015parallel}. For long accumulation times, and thus sensitive MS, the increase in throughout by parallel accumulation is proportional to the number of ions accumulated in parallel, and we expect to gain $2 - 10$ fold, \fg{opportunities}e. \\

\subsection*{Higher-multiplexing for increased throughput}
Achieving high chromatographic resolution and quantifying 1000s of proteins requires an hour of LC-MS/MS time or more. Thus, to quantify the proteomes of thousands of single cells within hours, we need to quantify many cells per LC-MS/MS run. Such multiplexing can be achieved by isobaric chemical barcoding \citep{Pappin2004multiplexed}. These barcodes are chemically identical, but distinguishable by MS due to their different isotopic compositions. While there are a number of commercially-available barcodes sets, the largest such set only allows the simultaneous analysis of eleven different samples. However, larger sets of barcodes can be synthesized and indeed many colleagues are actively working on making  multiplex tags that would allow $30-100$ individual cells to be measured in parallel (\fg{opportunities}f). In creating new barcodes for single cell proteomics, there is an opportunity to solve an outstanding problem with the technology, coisolation: quantitative fragments from the currently-available barcodes are not peptide specific, allowing the signal to be polluted by fragments from different peptide species\citep{savitski2013measuring}. The barcode labeling chemistry would have to be changed such that the barcode to peptide bond is much stronger or the barcodes require less energy for fragmentation, allowing the quantification to be based on peptide-specific complement ions. 

\subsection*{Protein identification and data analysis}
While the analysis of single cell MS data can use tools developed for bulk MS, it raises specific challenges and opportunities. One specific challenge is peptide sequencing. At very low abundance, many peptides may not produce enough fragment ions to support confident identification. One solution  may be to include a carrier channel together with single-cells, each of which labeled with a unique tandem-mass-tag\citep{scopems2017}. With this strategy, the fragment ions supporting peptide identification are derived from peptides pooled across all samples, including the carrier channel\citep{scopems2017}. Combining this strategy with further increases in the number of  distinct tandem-mass-tags will further facilitate peptide identification and decrease the need for carrier cells.  Another approach is to use additional peptides characteristics, such as retention time or empirical spectra. These approaches can exploit the fact that single cell proteomics data will contain multiple similar runs that can be used, for example, for making spectral libraries. As we discussed already, quantifying 1000s of single cells in a reasonable time frame requires barcoding, which will benefit from new computational approaches to accurately merge quantitative data sets acquired separately. If single cell proteomes are quantified by shotgun MS, the proteins quantified in  different runs will overlap partially, with some proteins quantified across all MS runs, while others only in a subset of the runs. This partial overlap will result in missing data, i.e., the levels of some proteins will not be quantified in some single cells. Such missing data are pervasive with respect to single cell RNA-seq\citep{hicks2017missing}, and must be handled carefully in single cell proteomics to avoid artifacts.

\bigskip

\noindent {\bf Acknowledgments:} We thank B. Budnik, G. Harmange, E. Emmott, A. Ivanov, S. Semrau, and B. Karger for constructive feedback. This work was funded by startup funds from Northeastern University and a New Innovator Award from the NIGMS from the National Institutes of Health to N.S. under Award Number DP2GM123497.

\noindent {\bf Competing Interests:} The authors declare that they have no
competing financial interests.

\noindent {\bf Corresponding author:} N.S. (nslavov@alum.mit.edu)
 
\noindent {\bf Contributions:} H.S. and N.S. conceived and wrote the manuscript.\\

\printbibliography

\end{spacing}  
\end{document}